# Field Emission Characterization of MoS$_2$ Nanoflowers

**Filippo Giubileo[1,\*], Alessandro Grillo[2], Maurizio Passacantando[3,1], Francesca Urban[2,1], Laura Iemmo[2,1], Giuseppe Luongo[2,1], Aniello Pelella[2], Melanie Loveridge[4], Luca Lozzi[3] and Antonio Di Bartolomeo[2,\*]**

[1] CNR-SPIN Salerno, via Giovanni Paolo II n. 132, Fisciano 84084, Italy; filippo.giubileo@spin.cnr.it; (F.G.)

[2] Physics Department "E. R. Caianiello", University of Salerno, via Giovanni Paolo II n. 132, Fisciano 84084, Italy; agrillo@unisa.it (A.G.); furban@unisa.it (F.U.); liemmo@unisa.it (L.I.); giluongo@unisa.it (G.L.); aniello.pelella@gmail.com (A.P.); adibartolomeo@unisa.it (A.D.B)

[3] Department of Physical and Chemical Science, University of L'Aquila, and CNR-SPIN L'Aquila, via Vetoio, Coppito 67100, L'Aquila, Italy; maurizio.passacantando@aquila.infn.it (M.P.); luca.lozzi@aquila.infn.it

[4] WMG, International Manufacturing Centre, University of Warwick, Coventry, CV4 7AL, United Kingdom; M.Loveridge@warwick.ac.uk (M.L.)

\* Correspondence: filippo.giubileo@spin.cnr.it (F.G.); adibartolomeo@unisa.it (A.D.B)

**Abstract:** Nanostructured materials have wide potential applicability as field emitters due to their high aspect ratio. We hydrothermally synthesized MoS$_2$ nanoflowers on copper foil and characterized their field emission properties, by applying a tip-anode configuration in which a tungsten tip with curvature radius down to 30-100 nm has been used as the anode to measure local properties from small areas down to 1-100 $\mu m^2$. We demonstrate that MoS$_2$ nanoflowers can be competitive with other well-established field emitters. Indeed, we show that a stable field emission current can be measured, with a turn-on field as low as 12 V/$\mu m$, and a field enhancement factor up to 880 at 0.6 $\mu m$ cathode-anode separation distance.

## 1. Introduction

The layered materials of the transition-metal dichalcogenides (TMDs) family have attracted enormous attention for their physical and chemical properties [1,2]. TMDs have a 2H crystal structure and chemical composition MX$_2$, (where M is a transition metal atom and X is a chalcogen atom); M atoms are covalently bonded to six chalcogen atoms to form X–M–X sandwich layers. The bulk materials are the result of the stacking of monolayers by weak van der Waals forces. The transition metal (Mo, W, Ti, Nb, etc.) and the chalcogen (S, Se, Te) determine TMD properties such as the band gap, the electron mobility or the thermal and chemical stability. Consequently, it is possible to obtain metals (NbS$_2$, VSe$_2$), semimetals (WTe$_2$, TiSe$_2$), semiconductors (MoS$_2$, WS$_2$), and superconductors (NbSe$_2$, TaS$_2$). Moreover, physical properties may be layer-dependent. For instance, for MoS$_2$ there is a transition from an indirect bandgap (of 1.3 eV) in the bulk material to a direct gap (1.8 eV) in the monolayer [3]. TMDs materials have demonstrated their suitability for several applications, such as energy storage [4-5], lithium-ion batteries [6], field effect transistors [7-13], electrocatalysis [14,15], gas sensors [16,17], solar cells[18], memory devices [19,20], and photodetectors [20,21]. Moreover, TMD nanostructures have very different behaviors in terms of electronic, optical and chemical properties, depending on the morphology. These can be monolayers [7,22-24], nanoflakes [25], nanotubes [26,27] and nanoflowers [28-30]. In particular, MoS$_2$ nanoflowers (NFs) have been reported to have good lithium storage properties [30]. This can be exploited for high-performance anodes, as well as an efficient catalytic activity for hydrogen evolution reactions [31]. Few studies have also focused the attention on the field emission (FE) properties of MoS$_2$ NFs [32,33]. Indeed, almost all conducting or semiconducting nanostructures are good candidates as field emitters due to the high aspect ratio that locally favors electric field enhancement [34]. Examples largely investigated are single carbon nanotube (CNT) [35,36], CNT films [37-41], nanowires [42,43], nanoparticles [44,45], and graphene [45-48]. Conversely, FE properties of $MoS_2$ have not until now been largely investigated. Only a few studies report FE measurements on $MoS_2$ flakes [49-51], nanoflowers [32,33], and edge-terminated vertically aligned (ETVA) films [52].

In this paper we report a detailed characterization of the field emission properties of hydrothermally synthesized $MoS_2$ nanoflowers. The flower-like configuration provides a great number of nanoflakes with free open edges suitable for high current emission. Moreover, the use of a tip-shaped anode allow to collect current emitted from small areas (from 1 to 100 $\mu m^2$) with higher spatial resolution than a standard parallel plate set-up. We show that $MoS_2$ nanoflowers are suitable materials for easy-to-fabricate cold cathodes featuring turn-on field as low as 12 V/$\mu$m and a field enhancement factor up to 880 for a cathode-anode separation distance of 600 nm.

## 2. Materials and Methods

$MoS_2$ was synthesized using ammonium molybdate (($NH_4$)$_6Mo_7O_{24}\cdot4H_2O$) and thiourea ($CH_4N_2S$) (by Sigma Aldrich). A standard procedure was used whereby (see Figure 1a) 0.70 g of ammonium molybdate and 4.48 g of thiourea were dissolved in deionized water (70 ml). The solution was stirred until a clear solution was obtained. Following, the solution was transferred to a Parr 5500 hydrothermal reactor for 12 hours at 220 °C at a pressure of 40 bar. Finally, the precipitated product, $MoS_2$, was washed with water and ethanol and dried. A scanning electron microscope (SEM, LEO 1530, Zeiss, Oberkochen, Germany) was used to image the samples revealing several MoS2 flower-like nanostructures distributed on the surface (Figure 1b) with nanosheets dimensions that have been statistically estimated as 100–200 nm for height, and 5-10 nm for thickness.

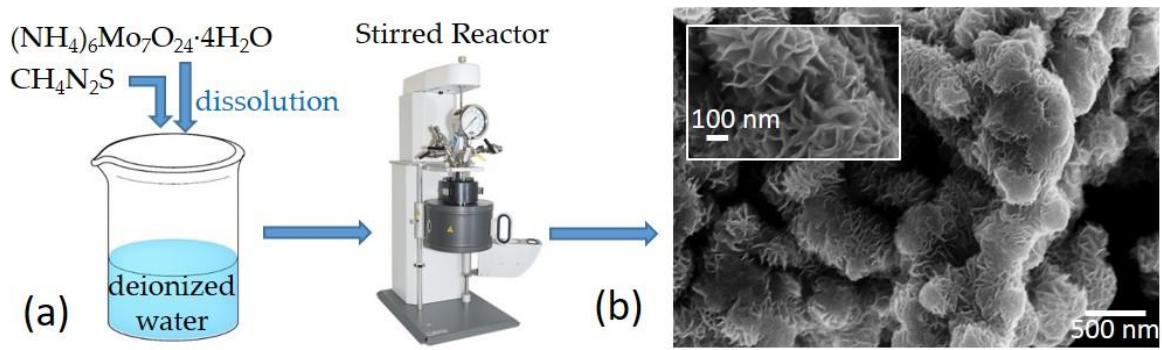

**Figure 1.** (a) Schematic of the $MoS_2$ synthesis by hydrothermal method. (b) Scanning electron microscope image of a $MoS_2$ nanoflower.

The surface elemental composition was analyzed by X-ray photoelectron spectroscopy (XPS). Measurements were taken using a PHI 1257 system equipped with a hemispherical analyzer and a non-monochromatic Mg $K_\alpha$ source (hν = 1253.6 eV) and operated with a base pressure in the chamber of $10^{-9}$ Torr.

Field emission measurements were performed inside the SEM chamber (at a pressure of $10^{-6}$ Torr and at room temperature) using two tungsten tips as electrodes (curvature radius of tip apex below 100 nm). These were mounted on piezo-driven nano-manipulators allowing fine movements with 5 nm step resolution. The electrical measurements were performed by means of a semiconductor parameter analyzer (Keithley 4200-SCS) in the bias range ±120V and with current resolution of about $10^{-14}$ A.

## 3. Results and Discussion

### 3.1. XPS Characterization

The surface chemical properties of MoS₂ nanoflowers were analyzed by XPS (results are shown in Figure 2). Spectra were acquired with a pass energy of 23.50 eV (overall experimental resolution of 0.8 eV), calibrated to the C 1s core level peak (284.8 eV) of the adventitious carbon [53] and fitted with Voigt profile on a Shirley-type background [54]. The calibration value to give the binding energies (BE) is obtained by the fit of the C1s spectrum (Figure 2a). C 1s core level spectra reveal several chemical states of carbon with a dominant lowest-BE peak due to carbon C−C and less intense contributions that appear at higher BE due to the presence of carbon functional groups [53]. The O 1 s fitted peaks located at 530.2, corresponding to the C=O and Mo−O bonds, 531.8 and 533.2 eV correspond to the energy of oxygen in C-O-C and OH−C bonds, respectively (Figure 2b).

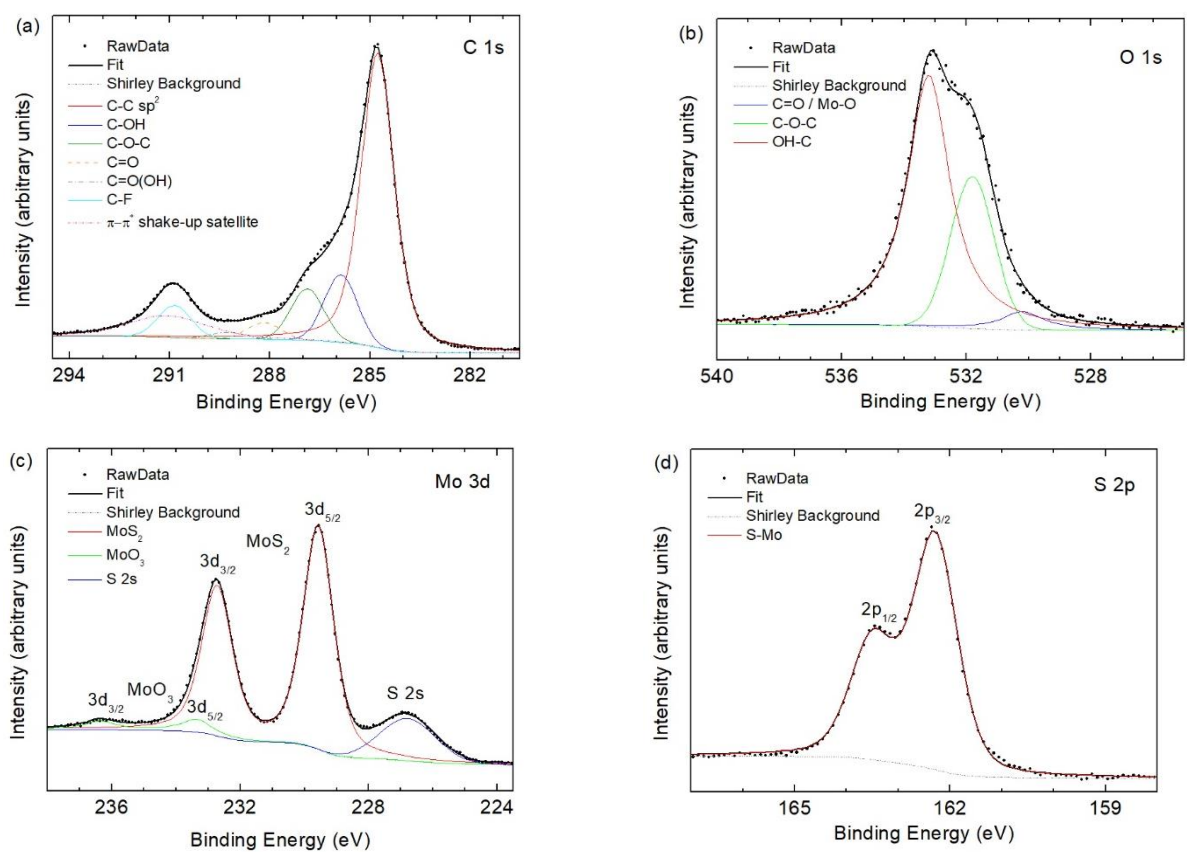

**Figure 2.** XPS Characterization of MoS₂ nanoflowers. Component peak-fitting of XPS spectra is shown for (a) C 1s region where C-C, C-OH, C-O-C, C=O, C=O(OH) and C-F are visible; (b) O 1s; (c) Mo 3d; (d) S 2p. Black solid lines represent the overall fit of experimental data (scattered points). The various peaks under the overall fitting curves represent the various components assumed to exist.

The high-resolution XPS spectra for Mo 3d and S 2p are shown in figures 2c and 2d, respectively. The Mo 3d spectrum (Figure 2c) exhibits two strong characteristic emission peaks at 232.7 (corresponding to Mo $3d_{3/2}$) and 229.6 (corresponding to Mo $3d_{5/2}$) eV. These BE values are consistent with electrons of $Mo^{4+}$ for MoS₂ [55]. Additionally, in this region, the S 2s peak at a binding energy of 226.6 eV corresponding to MoS₂ is also observed [55]. Another small contribution appears at binding energies of about 233 and 236 eV and it is attributed to the presence of MoO₃ [56,57]. Indeed, in the O 1s core-level spectrum, a weak peak at 530.9 eV shows a very small presence of MoO₃ ($O_2^{−}$ oxidation state). In Figure 2d the S 2p spectrum is shown, where the double peak corresponding S $2p_{1/2}$, and S $2p_{3/2}$ are clearly identified at 163.4 and 162.2 eV, respectively, with 1.2 eV spin–orbit energy separation corresponding to MoS$_2$ (S$_2-$ oxidation state) [58]. The XPS results are consistent with data already reported for MoS$_2$ single crystals, indicating that the nanosheets are in the semiconducting 2H phase [59]. From the XPS data it is possible to evaluate the elemental atomic concentration by:

$$C_x(\%) = \frac{I_x/S_x}{\sum_{i=1}^{n} I_i/S_i} \times 100$$

were $C_x$ is the atomic concentration of the $x$ element, $I_x$ is the peak area of the considered element and $S_x$ is the relative sensitivity factor. Therefore, considering the peak area in the XPS data of Figure 2 and taking into account the respective sensitivity factors $S_x$ for the 3d peak of Mo and the 2p peak of S that are 3.544 and 0.717 respectively, we obtained an atomic concentration of 32.7% and 67.3% for Mo and S respectively. Therefore, a stoichiometric ratio of Mo/S = 0.49 confirm the presence of the MoS$_2$ compound.

### 3.2. Field Emission Characterization

FE measurements on MoS$_2$ NFs were performed at room temperature inside the SEM chamber by contacting one of the two available W-tips directly on the sample surface (cathode) and positioning the second W-tip (anode) at a distance $d$ from the surface (Figure 3a). The cathode-anode separation distance $d$ can be precisely measured through SEM imaging by rotating the sample with respect the electron beam to have such distance almost perpendicular to the beam. Initially, both tungsten tips were approached on different areas of a MoS$_2$ NF (schematic is given as inset of Figure 3b) in order to measure a standard two-probe current-voltage (I-V) characteristic, to check the conductivity of the sample. We found linear ohmic behavior corresponding to a total resistance R$_{Tot}$ of about 90 kΩ (Figure 3b). For comparison, we repeated the measurements using Au-tips and we found once again a linear ohmic behavior with R$_{Tot}$≈6 kΩ. For the FE experiment, we chose the W-tips because of the availability of tip apex with very small radius of curvature (~100 nm). Indeed, it has been demonstrated [60] that the use of tip-shaped anode allows to extract FE current from small areas (down to 1μm$^2$ and less) depending on the tip curvature radius and on the cathode-anode separation distance. Consequently, the tip-shaped anode setup allows to obtain local information about the FE properties with respect to the standard parallel plate setup that typically probes areas of several mm$^2$. Moreover, smaller areas favor the possibility to probe the emitters with lower field amplification factor (<50) that are usually not detected because on large areas (of the order of mm$^2$) are hidden by often present protruding strong emitters with $\beta \sim 500\text{-}1000$.

In the following, FE characteristics will be analyzed in the framework of the Fowler-Nordheim (FN) theory [61] in which the dependence of the FE current $I$ on the applied bias voltage $V$ is expressed as:

$$I = A \cdot a \frac{\beta^2 V^2}{\varphi d^2} exp\left(-b \ d \frac{\varphi^{3/2}}{\beta V}\right) \tag{1}$$

where $A$ is the emitting area, $a = 1.54 \times 10^{-6} AV^{-2} eV$ and $b = 6.83 \times 10^9 eV^{-3/2} m^{-1} V$ are constants, $\varphi$ is the workfunction of the emitting surface (we assume $\varphi$ = 5.25 eV [49]) and $\beta$ is the field enhancement factor that takes into account the electric field amplification at the tip apex of the emitter. Accordingly to the model, a linear FN-plot, i.e. $ln(I/V^2)$ $vs$ $1/V$, with slope $m = -(bd\varphi^{3/2})/\beta$ is expected if the measured current is due to the FE phenomenon. Despite this model has been derived considering the electronic emission achieved from a flat metallic surface through a triangular potential barrier at zero kelvin, FN theory has proven to be a valid model to achieve a first-approximation understanding of the emission phenomena from several nanostructures. The model is still widely used today, although corrections would be required to take into account effects of non-zero temperature, series resistance, inhomogeneous work-functions, extreme curvatures, and different dimensionality of the emitters [62-67].

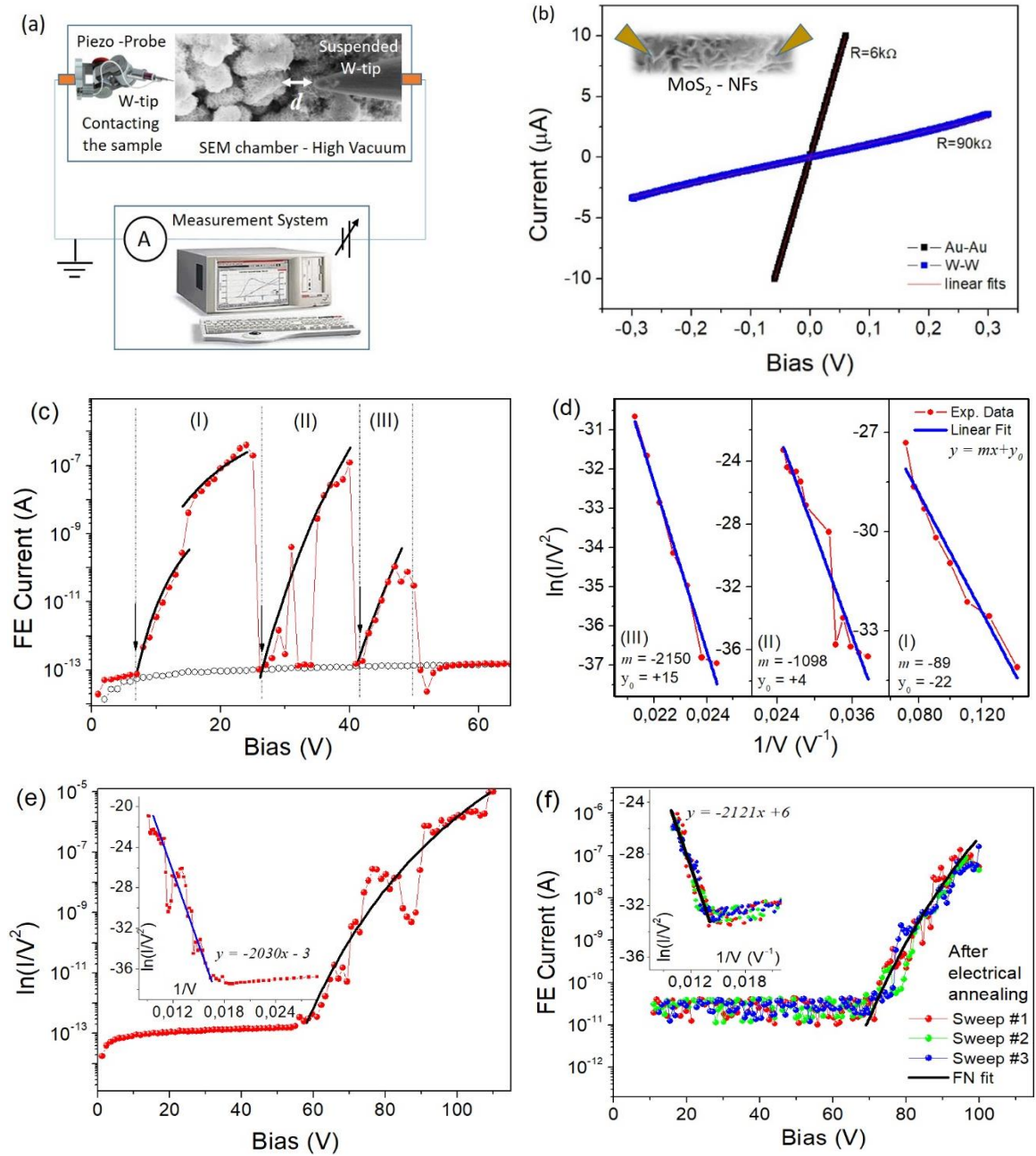

**Figure 3.** (a) Schematic of the FE measurement setup; (b) I-V characteristics measured by contacting both electrodes on the sample surface; (c) FE curve measured as first voltage sweep in a virgin area of the sample. Three successive regions of emission are identified. Black arrows indicate the turn-on voltage for each region. Black empty circles are experimental data measured in open circuit configuration. Black solid lines are the numerical simulations according to FN-theory (Equation 1); (d) FN-plots for the different regions. Solid lines are the linear fittings; (e) FE curve measured in a different location after the initial electrical stress; (f) Three consecutive voltage sweeps measured in a third different location always after electrical stress to show the FE repeatability.

To perform the FE measurements, one of the W-tips is retracted from the surface to act as the anode and the separation distance $d$ is finely tuned by means of the piezo-controlled nano-manipulators. The I-V characteristics are obtained by sweeping the voltage bias applied on the suspended W-tip (anode) from 0 to a maximum value of 120 V (to prevent setup damages) and measuring the FE current. We systematically observed that the first I-V characteristic measured in a virgin area is always characterized by a significant instability. As example, we show in Figure 3c a curve measured at a separation distance $d = 600$ nm: We observe that the emission started at 7 V (turn-on field $E_{on} \approx 12$ V/$\mu$m) and rapidly increases for more than six orders of magnitude, in a bias window of about 20 V. In this bias range, there is also an abrupt current raise of about one order of magnitude observed at 15 V. At an applied bias of 26 V there is a first sudden drop of the emitted current (from the maximum current $4.5 \cdot 10^{-7}$ A) to the floor noise (~$10^{-13}$ A) of the system (black empty circles in figure 3c represent the experimental data measured in open circuit configuration, i.e. having the suspended W-tip far away from the surface).

By continuing the raising voltage sweep, a second turn-on is visible ($E_{on} \approx 45$ V/$\mu$m), with the emitted current raising in the voltage range 28-40V again for about six orders of magnitude up to a maximum current of about $1.3 \cdot 10^{-7}$ A before a second sudden drop to the floor noise with no current emission. Finally, a third emission region is obtained for the bias range between 40 V and 50 V ($E_{on} \approx 67$ V/$\mu$m). In this case, lower current ($1.4 \cdot 10^{-10}$ A) is reached before the drop. For all these three regions (identified in Figure 3c) we compare the experimental results (colored scattered points) with the FN theoretical expectation from Equation 1 (black solid lines) to confirm the FE nature of the measured current. In Figure 3d we show the FN-plots corresponding to the three different regions indicated in Figure 3c. All FN-plots have a clear linear behaviour (solid lines represent the numerical linear fitting of the experimental data). The electrical conditioning of the sample surface, as described above, is a standard procedure used to stabilize the behaviour and the performance of large area emitters [40,60,68,69], in order to modify the surface towards a more homogeneous configuration that allows repeatability of the measurements. The observed features with several turn-on in the same voltage bias sweep is explained in terms of non-uniform array of emitters, with few protruding $MoS_2$ platelets with respect the multitude of the nanoflower. Indeed, the screening effect in non-uniform arrays may cause many emitters to become idle emitters, while few protruding platelets are overloaded by currying all current. Consequently, such emitters can burn out and/or evaporate by Joule heating [68, 69]. However, the overall structure of NFs remains unchanged, no variations being visible by SEM imaging. We notice that when a protruding emitter is burned by the too high current density, new emitters (previously idle) become active. However, due to a larger separation from the anode, they need a higher turn-on voltage to start the emission. From the linear fitting of the FN plots we can also estimate the field enhancement factor for the three different regions. As expected, the larger value is obtained for the more protruding flake, with $\beta \approx 550$. For a more accurate estimation, we have to take into account also the correction factor due to the tip-shaped anode setup. In such a case, a correction factor $k_{eff} = 1.6$ has to be considered [60] in the expression that relates the slope of the FN plot and the field enhancement factor $m = -(k_{eff} b d \varphi^{3/2})/\beta$, obtaining $\beta \approx 880$. For the other voltage regions, (II) and (III), we found $\beta \approx 72$ and $\beta \approx 37$, respectively. The lower field enhancement factor is responsible for the larger turn-on field requested to start the emission and it is explainable in terms of enhanced electrostatic screening effect provoked by the proximity of platelets to each other.

In Figure 3e we show a FE characteristic measured in the bias range 0-120 V after the electrical conditioning of the surface. The emission never drops to the noise floor, despite some fluctuations are still present, probably due to the desorption of adsorbed species, caused by sample heating and/or stretching and re-orientation of platelets. In particular, adsorbates [40,41] are usually present on the surface, originating regions with reduced work-function (and increased enhancement factor) that can cause the FE current instabilities. The observed FE current drops (of about one order of magnitude) are probably due to the evaporation of adsorbates from the sample surface, being more evident at larger current because caused by the local increase of temperature. However, few successive electrical sweeps are usually effective to stabilize the FE characteristics as demonstrated in Figure 3f in a different sample area. Reporting three successive voltage sweeps clearly demonstrates reproducibility and is in agreement with the expected FN behavior (solid line). Interestingly, for all curves reported in Figure 3e and 3f we found that the field enhancement factor is $\beta \approx 38$ and $\beta \approx 37$, respectively, the cathode-anode separation distance being always 600 nm. These data demonstrate that the FE characteristics measured on small areas of $MoS_2$ NFs are clearly reproducible after the electrical conditioning. In Table 1 we summarized the values of the slope $m$ resulting by the linear fitting of the FN-plots for experimental data of Figure 3 and the consequent extracted β values, the separation distance $d$ being fixed at d = 600 nm.

| Data | Slope $m$ of linear FN plot | Field enhancement factor β |
|---|---|---|
| Fig 3c Region(I) | -89 | 880 |
| Fig 3c Region (II) | -1098 | 72 |
| Fig 3c Region (III) | -2150 | 37 |
| Fig 3e | -2030 | 38 |
| Fig 3f | -2121 | 37 |

**Table 1.** Summary of FE parameters for curves of Figure 3.

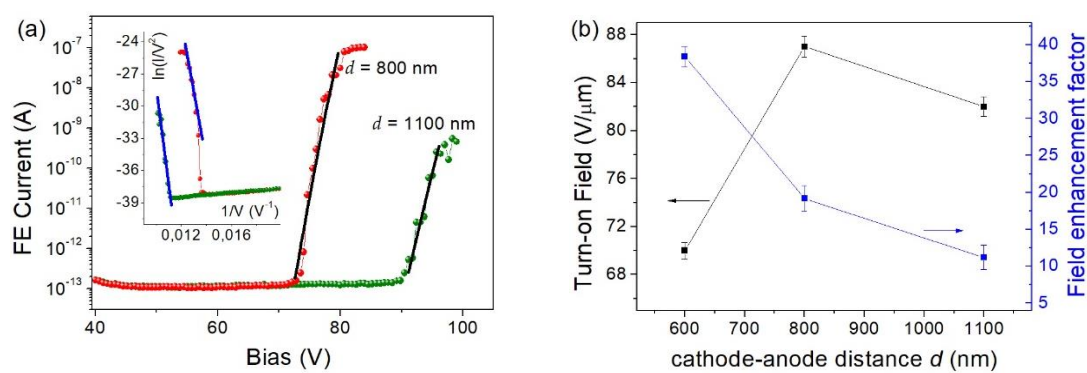

**Figure 4.** Effect of cathode-anode separation distance variation on the FE I-V characteristics. (a) Curves are measured for d = 800 nm and 1100 nm and are compared to theoretical FN behavior (solid lines). Inset: FN-plots and linear fittings. (b) Dependence of the turn-on field and of the field enhancement factor on the cathode-anode separation distance $d$ in the range 600 nm – 1100 nm.

Finally, we verified the effect of tuning the separation distance on the FE characteristics. In Figure 4a we show the I-V curves measured by increasing the distance $d$ to 800 nm and to 1100 nm. The numerical fittings (solid lines) confirmed that experimental data are a signature of field-emitted current according to the FN model, as also demonstrated by the linear behaviors of the corresponding FN plots in the inset. We established that at these distances a turn-on field above 80 V/μm is necessary to extract the current from the NFs while the field enhancement factor shows a decreasing trend for increasing distance, so that β ≈ 11 at $d$ = 1100 nm.

## 4. Conclusions

We performed field emission characterization of hydrothermally synthesized $MoS_2$ nanoflowers, with sheets having typical size between 100–200 nm, and thickness 5-10 nm. Using a tip-shaped anode setup, we demonstrate that small areas, down to 1 μm², can be probed, evidencing the presence of few protruding strong emitters characterized by field enhancement factor up to 880 that allow emission at turn-on field as small as 12 V/μm. After the electrical annealing that burns out the strong emitters, idle emitters become active, but due to the electric screening effect, larger field is necessary to extract electrons from the $MoS_2$ nanosheets. We demonstrate that the flower-like configuration provides a great number of nanoflakes with free open edges suitable for high current emission opening the technological scenario to realize large area emitting cold cathodes.